\documentclass[preprint,aip,jcp,showpacs]{revtex4-1}

\usepackage{color}
\usepackage{graphicx,amssymb,epsfig,epsf,amsmath}

\begin{document}

\title{Pair-correlation properties and momentum distribution of
 finite number of interacting trapped bosons in three dimension} 

\author{Anindya Biswas$^{1,}$\footnotemark[1] \footnotetext[1]{Corresponding author e-mail :
abc.anindya@gmail.com}, Barnali Chakrabarti$^{2,3}$ and 
Tapan Kumar Das$^{1}$}

\affiliation{$^{1}$Department of Physics, University of Calcutta, 
92 A.P.C. Road, Kolkata 700009, India \\
$^{2}$The Abdus Salam International Centre for Theoretical Physics,
 34100, Trieste, Italy\\
$^{3}$Department of Physics, Lady 
Brabourne College, P-1/2 Suhrawardy Avenue, Kolkata 700017, India}

\begin{abstract} 
We study the ground state pair-correlation properties of a weakly
interacting trapped Bose gas in three dimension by using a correlated many-body method. Use of the van der Waals 
interaction
potential and an external trapping potential shows realistic features. We also test the validity of shape-independent 
approximation in the calculation of correlation properties.
\vskip 5pt \noindent 
\end{abstract} 
\maketitle
\section{Introduction}
The last two decades have been witness to intense activities in experimental and
theoretical study of correlation properties of interacting quantum
systems. The problem is still
challenging and open in the quantum many-body systems which are
 non-integrable.  The task becomes simplified for integrable systems like one
dimensional uniform Bose gas described by the Lieb-Liniger (LL) model [1,2],
which assumes that particles interact via a $\delta$-function repulsive
potential.  Since the experimental observation of Bose Einstein condensate (BEC) 
in ultracold trapped alkali atomic vapors, a lot of theoretical and
experimental work has been done to study its correlation properties [3-12].
In the recent experimental situation, it is easy to achieve
a quasi-one-dimensional strongly interacting degenerate Bose gas in a highly
anisotropic trap [13-15] which is correctly described by the Lieb-Liniger model~[16]. 
The complementary case is the weakly interacting (via finite range forces) trapped 
Bose gas. Although it is commonly believed that the Gross Pitaveskii (GP) 
equation is adequate for weakly interacting Bose gases, but a more
rigorous and accurate many-body treatment, incorporating realistic
interatomic interactions and interatomic correlations, is crucial for
studying correlation properties in realistic condensates.
The first motivation of the present work is to investigate the 
importance of finite size and trapping 
effects on the ground state correlation properties of ultracold atomic BECs.
 The second motivation of our study is to assess the 
validity of shape-independent approximation in the correlation properties.
 Dilute BECs are known to possess shape-independent (SI) 
property which is frequently described by the $\delta$-function potential, 
whose strength is proportional to the $s$-wave scattering length $(a_{s})$. In the mean-field 
description, 
the effective interaction is determined by the factor $Na_{s}$ (where $a_s$ is 
expressed in units of oscillator length of the trap), whereas 
in the full many-body description we solve the many-body Schr\"odinger equation 
for different two-body potentials that generate identical $a_{s}$. 
Thus, our present calculation serves as a stringent test of the 
SI approximation and verifies whether the long range correlation
 at all affects the correlation properties or not. Inclusion of all 
possible two-body correlations in our {\it{ab initio}} many-body method correctly
 calculates correlation effects at zero temperature. 
Use of the realistic van der Waals interatomic potential with a long attractive tail 
provides the realistic aspects of the correlation function and its momentum 
distribution. Specially the effect of short range 
repulsion in the interatomic potential 
is expected to be reflected in the pair correlation function. \\

The paper is organized as follows. In the next section, we briefly review 
our theoretical approach. In Section III, we present the results of our 
calculation of one-body density, pair distribution function and their 
momentum distributions. Finally, in Section IV we draw our conclusions.\\

\section{Potential Harmonic Expansion Method}
We adopt the potential harmonics expansion method together with a short range 
correlation function (CPHEM), which has already been established as a successful 
and useful technique for investigating BEC with realistic two-body interactions. 
In the following, we briefly describe the technique. Interested readers can find 
the details in Refs.~\cite{das1,kundu,olp}. \\

We start with the many-body Schr\"odinger equation of $A$ spinless bosons trapped in an isotropic 
harmonic oscillator potential of frequency $\omega$ at zero temperature.
The center of mass
motion can be  separated by introducing the center of mass vector ($\vec{R}$) and 
$N=A-1$ Jacobi vectors $\{\zeta_1,...,\zeta_N\}$ defined as~\cite{ballot}
\begin{equation}
 \vec{\zeta}_{i}=\sqrt{\frac{2i}{i+1}}\left( \vec{x}_{i+1} - \frac{1}{i} 
\sum_{j=1}^{i} \vec{x}_{j}\right) , \hspace*{0.5cm} (i=1,...,N) ,
\end{equation}
where $\vec{x}_i$ is the position vector of the $i$-th particle. The relative 
motion of the system is described by
\begin{eqnarray} 
\Big[&-&\frac{\hbar^{2}}{m} \sum_{i=1}^{N} \nabla_{\zeta_{i}}^{2}+
V_{trap}(\vec{\zeta}_1, ..., \vec{\zeta}_N) + 
\nonumber
\\
&&V_{int}(\vec{\zeta}_{1}, ..., \vec{\zeta}_{N})-E_{R} \Big] 
\psi(\vec{\zeta}_{1}, ..., \vec{\zeta}_{N}) = 0 ,
\label{SE_rel_motion}
\end{eqnarray} 
where $V_{trap}$ and $V_{int}$ are respectively the trapping and pair-wise 
interaction potentials, expressed in terms of the Jacobi vectors. The energy of 
the relative motion is $E_R$. Next hyperspherical variables are introduced by 
defining a `hyperradius'
\begin{equation}
r=\left[\sum_{i=1}^N \zeta_i^2\right]^{\frac{1}{2}},
\end{equation}
and a set of $(3N-1)$ `hyperangles', consisting of $2N$ polar angles of $N$ 
Jacobi vectors and $(N-1)$ angles defining their {\it relative lengths}~\cite{ballot}. 
In the hyperspherical harmonics expansion method 
(HHEM) the $N$-body Schr\"odinger 
Equation~(\ref{SE_rel_motion}) is solved by expanding  $\psi$ in the complete 
set of hyperspherical harmonics (HH), which are the eigenfunctions of the 
$N$-dimensional Laplace operator (analogous to the spherical harmonics in 
three dimension)~\cite{ballot}. Substitution of this in Eq.~(\ref{SE_rel_motion}) and 
projection on a particular HH results in a set of coupled differential 
equations. However imposition of symmetry of the wave function and calculation 
of the matrix elements are such formidable tasks that a practical solution 
for $A>3$ is nearly impossible. Moreover due to the fact that the  
degeneracy of the HH basis increases very rapidly with the increase in the 
grand orbital quantum number 
$K$~\cite{ballot}, a convergent calculation using  
HHEM with {\it the full} HH basis is extremely computer intensive. This is 
the price one pays for keeping {\it all many-body correlations}. On the 
other hand, a typical experimental BEC is designed to be extremely 
dilute to eliminate the 
possibility of molecule formation through three-body collisions.  
Hence we assume that the probability of three and more particles 
to come within the range of interatomic interaction is negligible 
and the effect of the two-body correlations will be adequate 
for the full many-body wave function. Moreover, only two-body interactions 
are relevant. Therefore,  
$\psi$ can be expressed in terms of Faddeev components $\psi_{ij}$ of the 
$(ij)$ interacting pair~\cite{Fabre} 
\begin{equation}
\psi=\sum_{i<j}\psi_{ij}(\vec{r}_{ij},r).
\end{equation}
Since only two-body correlations 
are important, 
$\psi_{ij}$ is a function of the 
interacting-pair separation $r_{ij}=\vec{x}_i-\vec{x}_j$ and $r$ only. 
Hence it can be expanded in the subset of HH needed for the expansion of 
the two-body interaction $V(\vec{r}_{ij})$. This subset is called the 
potential harmonics (PH) basis~\cite{Fabre}. This results in a dramatic 
simplification in the analytic and computation works and it has 
been  used upto 14000 bosons in the trap~\cite{kundu}. In this procedure, 
a realistic interatomic interaction can be used. Such an interaction 
has a strongly repulsive core. Hence $\psi_{ij}$ must be vanishingly small 
for small values of $r_{ij}$. But the first ($K=0$) term of the PH 
basis is a constant~\cite{ballot}. Hence to reproduce the short-range behaviour of 
$\psi_{ij}$ correctly, a large number of PH basis functions is needed, 
which slows down the rate of convergence considerably. Hence, we 
introduce a `short-range correlation function' in the PH expansion basis~\cite{CDLin}. 
In the zero temperature BEC, the kinetic energy of the interacting pair 
is practically zero. Therefore, the two-body collision process is determined 
entirely by the $s$-wave scattering length ($a_s$). The small $r_{ij}$ 
behavior of $\psi_{ij}$ will be that of the zero-energy wave function 
$\eta(r_{ij})$ of the pair interacting via $V(r_{ij})$. Hence we 
introduce $\eta(r_{ij})$ as a short-range correlation function in the 
PH expansion~\cite{olp}. We have checked explicitly that this improves the rate 
of convergence dramatically. In order that the small $r_{ij}$ behavior of 
$\psi_{ij}(r_{ij},r)$ corresponds to the correct two-body interaction appropriate 
for the experimental $a_s$, we adjust the very short-range behavior of the 
interatomic interaction (in the case of 
the van der Waals potential used by us, it will be the hard-core radius $r_c$) 
so that $\eta(r_{ij})$ asymptotically becomes proportional to 
$(1-\dfrac{a_s}{r_{ij}})$~\cite{pethick}. 
Substitution of this expansion of $\psi$ in 
Eq.~(\ref{SE_rel_motion}) and projection on a particular PH gives rise to 
a set of coupled differential equations in $r$~\cite{olp}
\begin{eqnarray}
\Big[&-&\dfrac{\hbar^{2}}{m} \dfrac{d^{2}}{dr^{2}} + \dfrac{\hbar^{2}}{mr^{2}}
\{ {\cal L}({\cal L}+1) + 4K(K+\alpha+\beta+1)\} 
\nonumber 
\\
&+& V_{trap}(r) -E_R \Big] U_{Kl}(r) 
\\ 
&+& \sum_{K^{\prime}}f_{Kl}V_{KK^{\prime}}(r)f_{K'l} U_{K^{\prime}l}(r) = 0 ,
\nonumber
\label{FinalCDE}
\end{eqnarray}
where ${\cal L}=l+(3A-6)/2$, 
$U_{Kl}(r)=f_{Kl}u_{K}^{l}(r)$, $\alpha=\frac{3A-8}{2}$ and 
$\beta=l+\frac{1}{2}$.
Here, $l$ is the orbital angular momentum of the system (assumed to be contributed 
by the interacting pair only) and $u_K^l(r)$ is the coefficient of expansion 
of $\psi_{ij}(r_{ij},r)$ in the correlated PH basis, while $f_{Kl}$ 
represents the overlap of the PH corresponding to the ($ij$)-partition 
with the sum of PHs of all partitions~\cite{kundu}. Expressions 
for the potential matrix element $V_{KK^{\prime}}(r)$ and $f_{Kl}$ 
can be found in 
Ref.~\cite{kundu,olp}. \\

Eq.~(5) is solved by hyperspherical adiabatic approximation 
(HAA)~\cite{das3}, for which we assume that the hyperangular motion 
is much faster than the hyperradial motion. Consequently, the former can be 
separated adiabatically and solved to obtain an effective potential as a 
parametric function of $r$, in which the hyperradial motion takes place. 
The hyperangular motion is effectively solved by diagonalizing the potential 
matrix $V_{KK^{\prime}}(r)$ together with the hyper-centrifugal potential of 
Eq.~(5) to get the lowest eigenvalue $\omega_0(r)$ [corresponding 
eigen column vector being $\chi_{K0}(r)$], which is the effective potential 
for the hyperradial motion. Finally, the adiabatically separated 
hyperradial equation 
\begin{equation}
\Big[-\dfrac{\hbar^2}{m}\dfrac{d^2}{dr^2}+\omega_0(r)+
\sum_K|\dfrac{\chi_{K0}(r)}{dr}|^2-E_R\Big]\zeta_0(r)=0.
\end{equation}
is solved subject to appropriate boundary conditions to get $E_R$ 
and the hyperradial wave function $\zeta_0(r)$. The many-body wave 
function can be constructed in terms of $\zeta_0(r)$ and 
$\chi_{K0}(r)$~\cite{das3}.\\ 

\section{Results}
Now in order to study the correlation properties of the interacting 
Bose gas we choose a realistic interatomic potential having a strong repulsive core 
at a small separation and an attractive tail at large atomic separations.
 This is approximately represented by the 
van der Waals potential with a hard core of radius $r_{c}$ and a $\dfrac{1}{r^{6}}$ attractive tail, 
{\it viz.} $V(r_{ij})$ = $\infty$, for $r_{ij} \leq r_{c}$ and 
$-\dfrac{C_{6}}{r_{ij}^{6}}$ for $r_{ij}>r_{c}$. The effective 
interaction is characterized by 
the $s$-wave scattering length ($a_{s}$), which depends strongly on 
$r_{c}$ \cite{pethick}. Usually the potentials are chosen to be purely attractive or purely 
repulsive according to whether 
$a_{s}$ is negative or positive respectively. In our many-body calculation, 
we solve the zero-energy two-body Schr\"odinger equation with 
$V(r_{ij})$ to obtain $a_{s}$~\cite{olp,pethick}. The value of $r_{c}$ is adjusted 
so that 
$a_{s}$ has the values corresponding to the JILA experiments~\cite{iju,roberts}.
A typical value of $r_c$ is of the order of $10^{-3}$ o.u. In atomic units 
this is a few tens of Bohr, 
which is larger than the atomic radius. Note that $r_c$ is  
expected to be larger than the atomic radius, as the van der Waals potential 
with the hard core of radius $r_c$ is the {\it effective potential} which 
produces the correct experimental {\it 
zero-energy} 
scattering cross-section (given by $a_s$).  
\\ 
\subsection{One-body density}
We define the one-body density, 
as the probability density of finding a particle at a distance 
$\vec{r}_{k}$ from the center of mass of the condensate
\begin{equation}
R_1(\vec{r}_k)=\int_{\tau'}|\psi|^2d\tau'
\end{equation}
where $\psi$ is the full many-body wave function and the
integral over the hypervolume $\tau^{\prime}$ excludes the variable
$\vec{r}_{k}$.  After a lengthy but straightforward calculation we arrive at a closed form given by \\ 
\begin{eqnarray}
R_{1}(\vec{r}_{k})=\sqrt{2}  \int_{0}^{\infty} \int_{-1}^{1} 
2^{\alpha} \left[\frac{1}{\pi^{3/2}} \frac{\Gamma\left((D-3)/2\right)}
{\Gamma\left((D-6)/2\right)}\right] \left[\zeta_{0}(r')\right]^{2}
\nonumber
\\
\sum_{KK'} \chi_{K0}(r') \chi_{K'0}(r') 
(f_{Kl}f_{K'l})^{-1}(h_{K}^{\alpha \beta} h_{K'}^{\alpha \beta})^{-1/2}  
P_{K}^{\alpha\beta}(z) 
\nonumber
\\
P_{K'}^{\alpha\beta}(z)  r'^{D-4} 
\sqrt{\frac{1+z}{2}} \left(\sqrt{\frac{1-z}{2}}\right)^{D-8}
\nonumber
\\
\left(\sqrt{r'^{2}+2r_{k}^{2}}\right)^{-(D-1)} dr' dz,~~~~~~~~~~~~
\end{eqnarray}
where $D=3A-3$ and $h_{K}^{\alpha \beta}$ is the norm of the Jacobi polynomial 
$P_{K}^{\alpha \beta}(z)$. \\

The one-body density contains information regarding one
particle aspect of the bosonic system. Although it is not directly
measurable but in the interferometry experiment one can indirectly explore
it~\cite{rit,njp}. In Fig.1 we present calculated one-body density as a function of
the distance from the trap centre for a repulsive interaction corresponding to 
$a_{s}$ = .00433 o.u. for $10000$ $^{87}$Rb atoms in the 
JILA trap~\cite{iju}. In our calculation, length and energy are measured in oscillator 
units (o.u.) of length
$\left( a_{ho}=\sqrt{\frac{\hbar}{m\omega}}\right)$ and energy ($\hbar\omega$) respectively. 
For comparison, we also 
include the mean-field results. The effect of interaction is revealed by the deviation 
from the Gaussian profile. To explore the effect of interaction, 
we also calculate one-body density for a smaller $a_{s}$ value 
{\it viz.}, $a_{s}= 2.09 \times 10^{-4}$ o.u. for the same number of particles. 
Since the trapped condensate is always stable even for a large $A$, we see
appreciable changes in $R_1(\vec{r}_k)$ as $a_s$ decreases. For small $a_s$, the
density distribution is sharper as the correlations induced by the interactions 
are weak, while for the larger $a_s$, the peak is
flatter with a larger width. 
In Fig.~2 we present calculated one-body density for an attractive BEC with 
$a_{s} = -1.836 \times 10^{-4}$~o.u for different number of $^{85}$Rb atoms 
in the JILA trap~\cite{roberts}. 
The effective interaction parameter is $\lambda=A|a_s|/a_{ho}$. 
For $A$= 100, $\lambda$ is small and the system exhibits weak one-body 
density which extends to the trap size. 
Increasing $A$ gradually, the net effective attraction becomes
strong and the density becomes sharply peaked at a smaller distance.
\begin{figure}[hbpt]
\vspace{-10pt}
\centerline{
\hspace{-3.3mm}
\rotatebox{270}{\epsfxsize=6cm\epsfbox{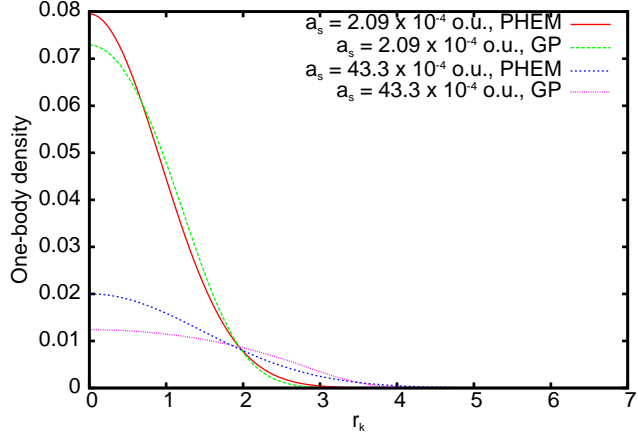}}}
\caption{(Color online) 
 One-body density distribution as a function of $r_k$ (in o.u.) for a repulsive BEC 
with A=10000 bosons. 
The choice of $a_{s}$ = 0.00433 o.u. corresponds to 
$^{87}$Rb experiment in the JILA trap. PHEM corresponds to our
 present many-body results and GP corresponds to mean-field results.}
\end{figure}
\begin{figure}[hbpt]
\vspace{-10pt}
\centerline{
\hspace{-3.3mm}
\rotatebox{270}{\epsfxsize=6cm\epsfbox{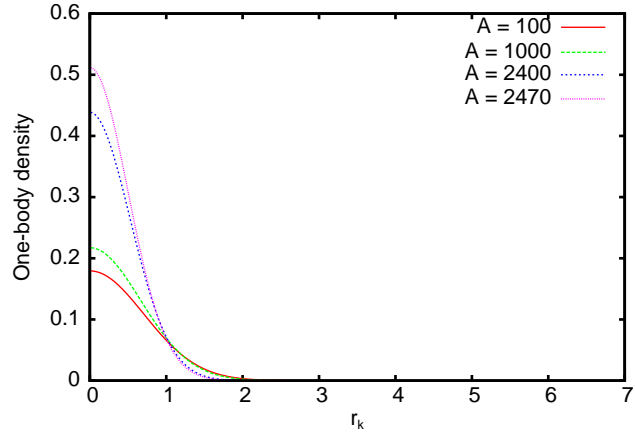}}} 
\caption{(Color online)
One-body density distribution as a function of $r_k$ (in o.u.) 
for an attractive
interaction ($a_s=- 1.836 \times 
10^{-4}$ o.u. for $^{85}$Rb atoms in the JILA trap), for various
 indicated values of particle numbers.} 
\end{figure}
\\
Taking the Fourier transform of $R_{1}(\vec{r}_{k})$, we obtain the one-body
momentum distribution.  One-body density is an abstract
concept. However, its fourier transform gives the experimentally measurable
quantity, the momentum distribution.  In Fig.~3, we plot our results for repulsive
interactions. It evolves from a Gaussian in the noninteracting limit to a curve having a 
sharper peak, as the net interaction increases. The peak at $k=0$ becomes 
more pronounced with increase in
effective repulsion, whereas for weak interaction it develops a long-range
tail in the momentum space. The momentum is being redistributed to higher $k$ values.  
The width of the 
low-momentum peak for $a_{s}=0.00433$ o.u. and $A=10000$ is about $0.7$ $\mu$m$^{-1}$.
\begin{figure}[hbpt]
\vspace{-10pt}
\centerline{
\hspace{-3.3mm}
\rotatebox{0}{\epsfxsize=8cm\epsfbox{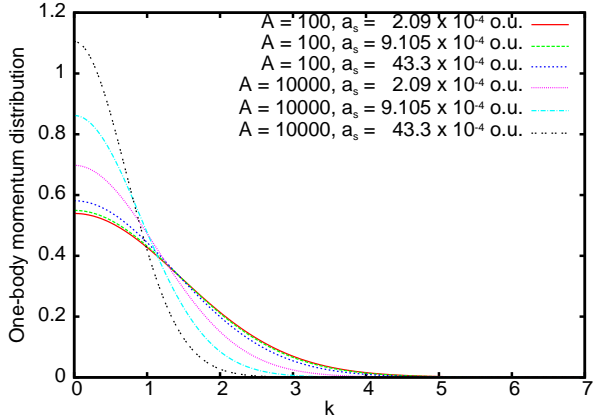}}}
\caption{(Color online) Calculated one-body momentum distribution for the repulsive
 BEC for different values of A and $a_{s}$. The choice of $a_{s}$ = 0.00433 o.u.
 corresponds to $^{87}$Rb experiment in JILA trap. All quantities are in 
appropriate oscillator units.}
\end{figure}

We have remarked  earlier that with recent progress in creating atomic clouds with 
large dipole moment, interest has been shifted to longer range interaction instead 
of taking only contact interaction. Thus, even in the low-density limit the use of 
a realistic interatomic interaction potential in the 
many-body calculation has been emphasized by several authors~[27-29]. The van der
 Waals potential is an ideal choice as it properly takes care of the effect of 
realistic dipole-dipole interaction. The strength of the van der Waals interaction 
$C_{6}$ changes widely from a small value for H atoms to a high value 
for Cs atoms. Hence, to see how the one-body density is affected by the strength of
 the long-range tail of the two-body potential, we select 
three different values of $C_{6}$ in addition to the actual experimental value -- 
one below and two above. These are (in o.u.): 
$5\times10^{-11}$, $6.489755 \times 10^{-11}$, $8.5\times 10^{-11}$, $8 \times 10^{-10}$, 
the second one being the experimental value. For each value of $C_{6}$, 
we calculate corresponding $r_{c}$ 
as before, such that $a_{s}$ has the experimental value 0.00433 o.u. For each set of
 $(C_{6}, r_{c})$, we solve the many-body 
Schr\"odinger equation as before. Calculated one-body density for different 
sets are shown in Fig. 4 for 10000 bosons. We observe that the 
one-body density of the inhomogeneous gas for 
different two-body potentials are almost indistinguishble; i.e. independent of the 
shape of $V(r_{ij})$. Our numerical calculations confirm that 
one-body density is absolutely determined (within numerical errors) by the 
parameter $a_{s}$ only.

\begin{figure}[hbpt]
\vspace{-10pt}
\centerline{
\hspace{-3.3mm}
\rotatebox{270}{\epsfxsize=6cm\epsfbox{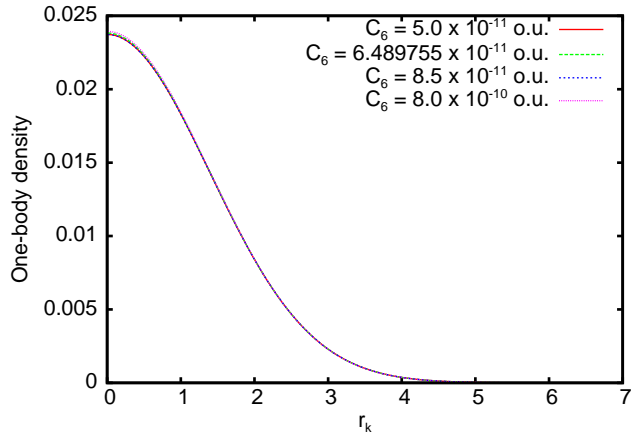}}}
\caption{(Color online) One-body density distribution for the repulsive BEC with
 several $C_{6}$ parameters, all of which 
correspond to the  identical $a_{s}$ = 0.00433 o.u., with 10000 $^{87}$Rb atoms 
in the JILA trap. 
All quantities are in oscillator units.}
\end{figure}
This is in contrast with ref.~\cite{vha}, where ground state energy of the 
condensate was found to depend on the shape of the potential. Thus the 
one-body density does not depend strongly on the shape of the interaction 
potential, while the ground state energy does.\\

An explanation of the above observation is as follows. The total condensate energy 
depends on the minimum of the well, in addition to the actual shape of 
the lowest eigen potential. 
In Table~1, we present the dependence 
of the position ($r_m$) and the value ($\omega_{0m}$) of the minimum of 
lowest eigen potential 
for different values of $C_6$ parameter, corresponding to the same $a_s$.\\[0.5cm] 
Table~1. Dependence of $r_m$ and $\omega_{0m}$ on $C_6$ (corresponding to 
$a_s=0.00433$) for $A=10000$ atoms 
(all quantities are expressed in o.u.).\\
\begin{center}
\begin{tabular}{|c|c|c|}
\hline
$C_6$ & $r_m$ & $\omega_{0m}$ \\
\hline
$5.0 \times 10^{-11}$ & $339.92$ & $48792.7$ \\
$6.4 \times 10^{-11}$ & $339.47$ & $48668.7$ \\
$8.5 \times 10^{-11}$ & $339.04$ & $48546.3$ \\
$8.0 \times 10^{-10}$ & $338.93$ & $48517.8$ \\
\hline
\end{tabular}
\end{center}
\vskip 15pt
We observe that, as $C_6$ increases, $\omega_{0m}$ decreases appreciably, 
while position of the minimum of effective many-body potential 
changes by a small amount. In the next sub-section, we will see that 
the shape of the $\omega_0(r)$ curve 
remains practically unchanged. 
As the total condensate 
energy depends on the depth of the potential, as well as on its shape (stiffness), 
we observe that the ground state energy decreases gradually with increase in 
$C_6$~\cite{vha}. Thus the shape independence hypothesis is violated for the 
total energy. However, the one-body density distribution is given by the 
many-body wave function, which is independent of the minimum of the potential, but 
depends only on its shape. Hence it remains unchanged with 
change in $C_6$ parameter and shape independence of the one-body density profile 
is satisfied. \\
\subsection{Pair distribution function}
Another key quantity is the pair distribution function
$R_{2}(r_{ij})$, which determines the probability of finding the $(ij)$-pair of
particles at a relative separation $r_{ij}$. 
The study of pair
correlation is important, since the interatomic interactions play a crucial
role as there are two competing interaction length scales. When the atoms
try to form clusters, a strong very short range repulsion in atomic
interaction comes into play. As atoms repel each other strongly at very small
separations it is impossible to get some non-vanishing value of $R_{2}(r_{ij})$ at
$r_{ij}$ = 0. We calculate it as
\begin{equation}
R_2(r_{ij})=\int_{\tau ''}|\psi|^2d\tau '',
\end{equation}
where $\psi$ is the many-body wavefunction and the
integral over the hypervolume $\tau ''$ excludes integration over ${r}_{ij}$. 
Again after a lengthy calculation we can put it in a closed 
form given by
\begin{eqnarray}
R_{2}(r_{ij})=\sqrt{2}\int_{-1}^{1}
\left(\frac{1-z}{2}\right)^{\alpha}\left(\zeta_{0}
\left(r_{ij}\sqrt{\frac{2}{1+z}}\right)\right)^{2}
\nonumber
\\
\sum_{KK'}\left(\frac{h_{K}^{\alpha\beta}}{2^{\alpha}}\right)^{-1/2}
\left(\frac{h_{K'}^{\alpha\beta}}{2^{\alpha}}\right)^{-1/2}\left(f_{Kl}f_{K'l}\right)^{-1}
\nonumber
\\
\chi_{K0}(r)\chi_{K'0}(r)
P_{K}^{\alpha\beta}(z)P_{K'}^{\alpha\beta}(z) dz.~~~~~
\end{eqnarray}
We plot $R_{2}(r_{ij})$ in Fig.~5 for the attractive interaction with
different particle numbers. As mentioned earlier, pair correlation vanishes as
 $r_{ij}$ $\rightarrow$ 0, due to the strong interatomic repulsion and it cannot 
extend  beyond the size of the condensate. Hence $R_{2}$ is peaked at some 
intermediate value of $r_{ij}$. For weak interactions (small
$\lambda$), the correlation length is large.  However as the
effective attractive interaction increases, two interacting particles come
closer and the pair-correlation length decreases; the pair distribution function
becomes sharply peaked. It indicates stronger pair correlation in the system,
in agreement with expectations.  However our results are at variance 
with those obtained in LL model~\cite{ast}, which describes uniform systems with no
confinement. Consequently pair correlation function approaches its maximum 
asymptotically, whereas in our confined three dimensional case, it vanishes
asymptotically.  When $A$ is increased to 2470, which is very close to the
critical number ( $A_{cr} \simeq$ 2475), the effect is quite prominant.
As the width of the curves decrease with increase in effective attractive
interaction, the interacting pair becomes more localized. This manifests the
 possiblity of clustering due to large
two-body interaction and three-body recombination, if the attraction
becomes strong enough. This will lead to an eventual collapse of the attractive
condensate, when the pair-correlation length will drastically reduce to
a very small value and the condensate will be destroyed.\\
\begin{figure}[hbpt]
\vspace{-10pt}
\centerline{
\hspace{-3.3mm}
\rotatebox{0}{\epsfxsize=8cm\epsfbox{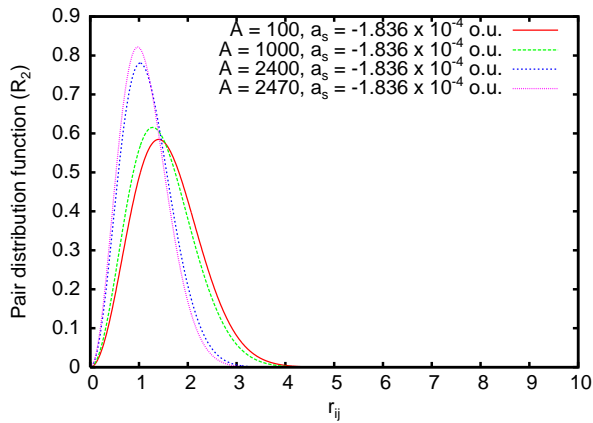}}}
\caption{(Color online) Plot of pair distribution function ($R_2(r_{ij})$) 
against $r_{ij}$ (in o.u.) for an attractive Bose 
gas with different particle numbers in the trap.}
\end{figure}

Two-particle correlation is directly related with pair inelastic processes 
and can be directly measured in photoassociation in interatomic
collisions. As the attractive interaction increases, the gas becomes highly
correlated and different inelastic processes may take place. The opposite case
is the repulsive Bose gas where the pair inelastic process will be
suppressed with increase in interaction. For completeness, in Fig.~6 we
plot the relative momentum distribution obtained as the Fourier transform of 
$R_{2}(r_{ij})$ for various particle numbers for
$^{87}$Rb condensate in the JILA trap, with $a_{s}$ = $0.00433$ o.u. ($100$ a.u.). The width
of the curve decreases and exhibits a sharp peak at $k = 0$, whose height
increases with increasing number of particles.\\ 

\begin{figure}[hbpt]
\vspace{-10pt}
\centerline{
\hspace{-3.3mm}
\rotatebox{0}{\epsfxsize=8cm\epsfbox{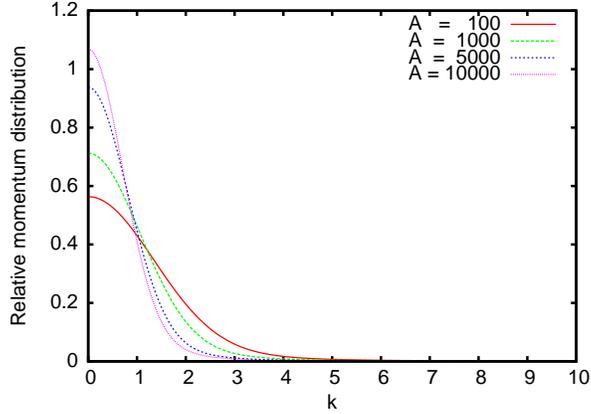}}}
\caption{(Color online) Plot of relative momentum distribution against k
 (in (o.u.)$^{-1}$)for various 
indicated values of particle numbers. 
The choice of $a_{s}$ = 0.00433 o.u. corresponds to $^{87}$Rb experiment
in JILA trap.}
\end{figure}
Lastly to visualize the effect of different strength parameter $C_{6}$ 
on the pair-correlation, we plot $R_{2}(r_{ij})$ against $r_{ij}$
for the previous sets of potential parameters in Fig.~7. We find that
 different interatomic potential shapes corresponding to the same $a_{s}$ produce 
identical correlation properties, which
indicates that the calculated property is independent of the 
shape of the two-body potential. The fact that the ground state energy is 
dependent on the shape of the two-body potential~\cite{vha} whereas 
the one-body density and 
the pair distribution function are not, can be attributed to the fact that the 
many-body wave function remains invariant irrespective of the variation of the 
shape of the two-body potential due to variation of $(C_6,r_c)$ for the same value 
of $a_s$. Note that the condensate wave function in the hyperradial space 
$\zeta_{0}(r)$ is obtained by solving the hyperradial equation~(6) in the 
effective many-body potential $\omega_{0}(r)$. The shape and stiffness of this effective 
potential remain unchanged 
for the same value of $a_{s}$, irrespective of the variation of the 
shape of the two-body potential. However, as discussed earlier, the position and 
value of the minimum of the effective many-body potential 
change with the variation of the parameter $C_{6}$. 
To demonstrate this, we plot the 
effective many-body potential $\omega_{0}(r)$ as a function of 
$r$ for various values of $C_{6}$
in Fig.~8. The curves have been shifted both vertically and horizontally in order that 
the minima coincide (both in position and value). One observes that all the curves 
overlap completely (within numerical errors) showing that the shape and stiffness of 
the effective potential remain invariant. Hence the kinetic energy and the 
wave function remain unchanged, while potential energy and the total energy 
change with change of $C_6$. By direct numerical calculation, we have 
checked that the change of kinetic energy is very small compared with that 
of the total energy, as $C_6$ varies.
Therefore the ground state energy 
is shape dependent, whereas the one-body density and pair distribution function are not. \\
\begin{figure}[hbpt]
\vspace{-10pt}
\centerline{
\hspace{-3.3mm}
\rotatebox{0}{\epsfxsize=9cm\epsfbox{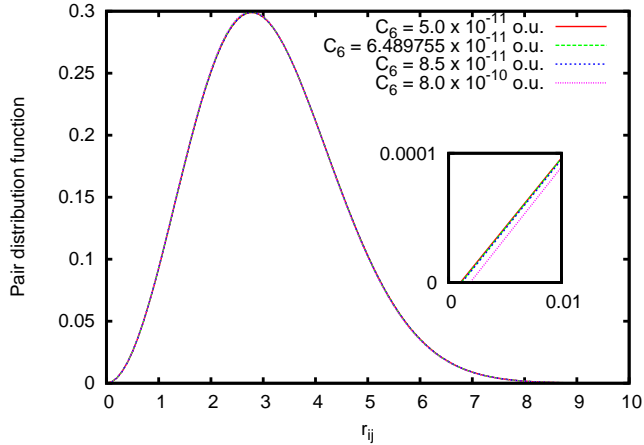}}}
\caption{(Color online) Plot of two-body correlation function against
 $r_{ij}$ in o.u. for repulsive BEC with several indicated $C_{6}$ parameters, all of which 
correspond to identical $a_{s}$ = 0.00433 o.u. with 10000 $^{87}$Rb atoms in JILA trap.
(The portion near the origin is magnified in the inset to show that the pair distribution function
 vanishes inside the hard core. The same is true for Fig.~5.)}
\end{figure}
\begin{figure}[hbpt]
\vspace{-10pt}
\centerline{
\hspace{-3.3mm}
\rotatebox{270}{\epsfxsize=6.7cm\epsfbox{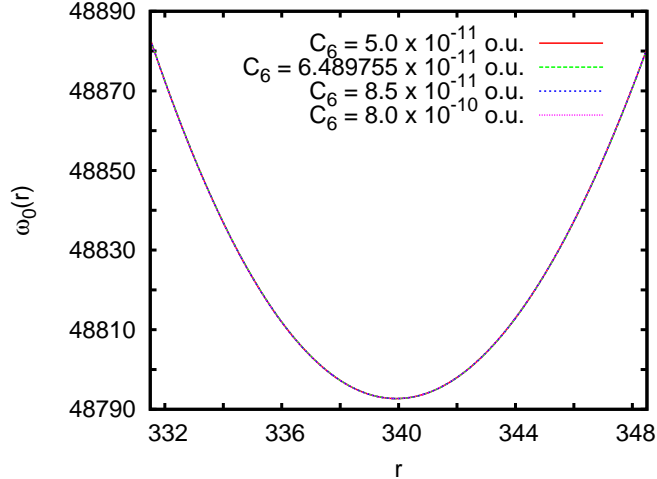}}}
\caption{(Color online) Plot of the effective potential (in o.u.) against $r$ (in o.u.) for 10000 $^{87}$Rb atoms
with $a_{s}$ = 0.00433 o.u.. The 
curve corresponding to $C_{6}=5.0\times~10^{-11}$ is in its actual position. The other curves have been shifted
both vertically and horizontally to coincide the minima (both in position and value). All 
the curves overlap completely (within numerical errors). This shows that the shape of the 
effective many-body potential $\omega_0(r)$ is independent of the choice of $(C_6,r_c)$ 
for the same value of $a_s$.}
\end{figure}

\section{Conclusion}
In conclusion, the present work focuses on the one-body density and 
pair-correlation aspects of the zero temperature
weakly interacting trapped Bose gas. The use of correlated many-body
approach takes care of the effect of finite-size, where quantum fluctuation is
 important and gives a realistic picture of correlation properties. 
Due to the use of a realistic interatomic interaction and consideration of
 finite number of atoms in the trap, 
our results deviate from the earlier results~\cite{nar,ast}, but exhibit realistic 
aspects which are relevant to experiments.
Our calculation is performed for a two-body potential (van der Waals potential),
 whose parameters can be adjusted to give both positive and negative scattering 
lengths, for Rb atoms in 
the JILA trap. Thus, our results are realistic and can be experimentally verified 
in future. Our calculations also verify the validity of the shape-independent 
approximation in dilute BECs for one-body density and pair-distribution functions.
 This is in contrast with the earlier observed shape dependence for ground state 
energies~\cite{vha}.\\
\begin{center}
\bf{Acknowledgement}
\end{center}
We would like to acknowledge 
helpful discussion with L. Salasnich, S.R. Jain and G. Mussardo. This work has been 
funded by the Department of Science and Technology (DST), India under grant number 
SR/S2/CMP/0059(2007). BC thanks the International Center for 
Theoretical Physics (ICTP), Trieste for hospitality and financial support, where a
 part of the work 
was done. TKD acknowledges the University Grants Commission (UGC), India for the
 Emeritus Fellowship. AB acknowledges the UGC, India for the Research Fellowship
 in Sciences for Meritorious Students and the Council of Scientific and Industrial 
Research (CSIR),  India for a Senior Research Fellowship.\\

\end{document}